\documentclass[twocolumn,preprintnumbers,amsmath,amssymb]{revtex4}

\usepackage{graphicx}% Include figure files
\usepackage{dcolumn}% Align table columns on decimal point
\usepackage{bm}% bold math

\begin{document}

%\preprint{PREPRINT L-2}

\title{Diffraction and an infrared finite gluon propagator} % Force line breaks with \\

\author{E.G.S. Luna$^{1,2}$}
%\altaffiliation[Also at ]{Physics Department, XYZ University.}%Lines break automatically
%or can be forced with \\
\affiliation{
$^{1}$Instituto de F\'{\i}sica Te\'orica,
UNESP, S\~ao Paulo State University, 01405-900, S\~ao Paulo, SP, Brazil \\
$^{2}$Instituto de F\'{\i}sica Gleb Wataghin,
Universidade Estadual de Campinas,
13083-970, Campinas, SP, Brazil}

\begin{abstract}
We discuss some phenomenological applications of an infrared finite gluon
propagator characterized by a dynamically generated gluon mass. In
particular we compute the effect of the dynamical gluon mass on $pp$ and
${\bar{p}}p$ diffractive scattering. We also show how the data on $\gamma p$ photoproduction
and hadronic $\gamma \gamma$ reactions can be derived from the $pp$ and $\bar{p}p$
forward scattering amplitudes by assuming vector meson dominance
and the additive quark model.

\end{abstract}


\maketitle

\section{A QCD-inspired eikonal model with a gluon dynamical mass}

Nowadays, several studies support the hypothesis that the gluon may develop a
dynamical mass \cite{cornwall,alkofer}.
This dynamical gluon mass, intrinsically related to an infrared finite gluon
propagator \cite{ans}, and whose existence is strongly supported by recent QCD lattice
simulations \cite{lqcd}), has
been adopted in many phenomenological studies \cite{halzen,natale01,natzan01}. Hence it
is natural to correlate the arbitrary mass scale that appears in
QCD-inspired models with the dynamical gluon one, obtained by Cornwall \cite{cornwall} by means of the
pinch technique in order to derive a gauge invariant Schwinger-Dyson equation for the gluon propagator.
This connection can be done building a QCD-based eikonal model where the onset of the dominance of
gluons in the interaction of high-energy hadrons is managed by the dynamical gluon mass scale.

A consistent calculation of high-energy hadron-hadron cross sections compatible with unitarity constraints
can be automatically satisfied by use of an eikonalized treatment
of the semihard parton processes. In an eikonal representation, the total cross sections is given by
\begin{eqnarray}
\sigma_{tot}(s) = 4\pi \int_{_{0}}^{^{\infty}} \!\! b\, db\, [1-e^{-\chi_{_{I}}(b,s)}\cos
\chi_{_{R}}(b,s)],
\label{degt1}
\end{eqnarray}
where $s$ is the square of the total center-of-mass energy and $\chi (b,s)$ is a complex
eikonal function: $\chi(b,s)=\chi_{_{R}}(b,s)+i\chi_{_{I}}(b,s)$. In
terms of the proton-proton ($pp$) and antiproton-proton
($\bar{p}p$) scatterings, this combination reads
$\chi_{pp}^{\bar{p}p}(b,s) = \chi^{+} (b,s) \pm \chi^{-} (b,s)$.
Following the Ref. \cite{luna2bjp}, we write the even eikonal as the sum of gluon-gluon, quark-gluon, and
quark-quark contributions:
\begin{eqnarray}
\chi^{+}(b,s) &=& \chi_{qq} (b,s) +\chi_{qg} (b,s) + \chi_{gg} (b,s) \nonumber \\
&=& i[\sigma_{qq}(s) W(b;\mu_{qq}) + \sigma_{qg}(s) W(b;\mu_{qg}) \nonumber \\
& & + \sigma_{gg}(s) W(b;\mu_{gg})] .
\label{final4}
\end{eqnarray}

\begin{figure}
\label{difdad01}
%\vspace{2.0cm}
\begin{center}
%\vspace{-0.6cm}
\includegraphics[height=.29\textheight]{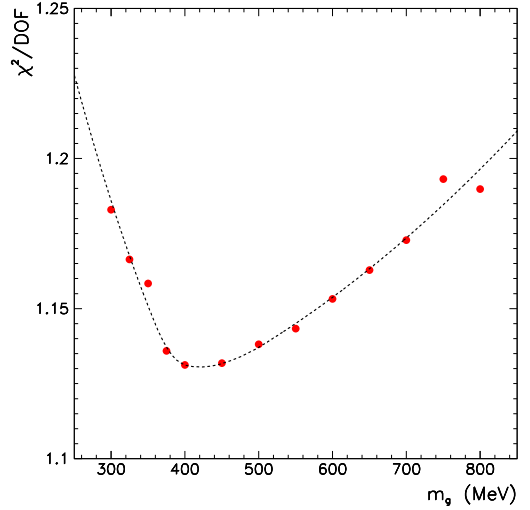}
\caption{The $\chi^{2}/DOF$ as a function of dynamical gluon mass $m_{g}$. }
\end{center}
\end{figure}

Here $W(b;\mu)$ is the overlap function at impact parameter space and $\sigma_{ij}(s)$ are the elementary
subprocess cross sections of colliding quarks and gluons ($i,j=q,g$). The overlap function is associated
with the Fourier transform of a dipole form factor,
$W(b;\mu) = (\mu^2/96\pi)\, (\mu b)^3 \, K_{3}(\mu b)$,
where $K_{3}(x)$ is the modified Bessel function of second kind. The odd eikonal $\chi^{-}(b,s)$, that
accounts for the difference between $pp$ and $\bar{p}p$ channels, is parametrized as
\begin{eqnarray}
\chi^{-} (b,s) = C^{-}\, \Sigma \, \frac{m_{g}}{\sqrt{s}} \, e^{i\pi /4}\, 
W(b;\mu^{-}),
\label{oddeik}
\end{eqnarray}
where $m_{g}$ is the dynamical gluon mass and the parameters $C^{-}$ and $\mu^{-}$ are constants to be
fitted. The factor $\Sigma$ is defined as $\Sigma \equiv 9\pi \bar{\alpha}_{s}^{2}(0)/m_{g}^{2}$,
with the dynamical coupling constant
$\bar{\alpha}_{s}$ set at its frozen infrared value. The eikonal
functions $\chi_{qq} (b,s)$ and
$\chi_{qg} (b,s)$, needed to describe the lower-energy forward data, are simply parametrized with terms
dictated by the Regge phenomenology \cite{luna2bjp}; the gluon term $\chi_{gg}(b,s)$, that
dominates the asymptotic behavior of hadron-hadron total cross sections,
is written as $\chi_{gg}(b,s)\equiv \sigma_{gg}(s)W(b; \mu_{gg})$, where
\begin{eqnarray}
\sigma_{gg}(s) = C^{\prime} \int_{4m_{g}^{2}/s}^{1} d\tau \,F_{gg}(\tau)\,
\hat{\sigma}_{gg} (\hat{s}) .
\label{sloh1}
\end{eqnarray}

Here $F_{gg}(\tau)\equiv [g \otimes g ](\tau)$ is the convoluted structure function for pair $gg$, $C^{\prime}$
is a normalization constant and $\hat{\sigma}_{gg}(\hat{s})$ is the subprocess cross section, calculated
using a
procedure dictated by the dynamical perturbation theory \cite{pagels}, where amplitudes that do not vanish to
all orders of perturbation theory are given by their free-field values, whereas amplitudes that
vanish in all orders in perturbation theory as $\propto \exp{(-1/4\pi \alpha_{s})}$ are retained at lowest order.
Only recently the physical meaning of the parameter $C^{\prime}$ has become fully \cite{natzan01}: it is
a normalization factor that appears in the gluon distribution function (at small $x$ and low $Q^{2}$) after the
resummation of soft emission in the leading $\ln (1/x)$ approximation of QCD,
\begin{eqnarray}
g(x) = C^{\prime}\, \frac{(1-x)^5}{x^{J}},
\label{distgf}
\end{eqnarray}
where $J$ controls the asymptotic behavior of $\sigma_{tot}(s)$. The results of global fits to
all high-energy
forward $pp$ and $\bar{p}p$ scattering data above $\sqrt{s}=10$ GeV and to the elastic differential scattering
cross section for $\bar{p}p$ at $\sqrt{s}=1.8$ TeV are shown in Figs. 1 e 2. The Figure 1
enables us to estimate a dynamical gluon mass $m_{g}\approx 400^{+350}_{-100}$ MeV. More details of the
fit results can be seen in Ref. \cite{luna2bjp}. The results of the fits to $\sigma_{tot}$ for both $pp$ and
$\bar{p}p$ channels are displayed in Fig. 2 in the case of a dynamical gluon mass $m_g = 400$ MeV, which is
the preferred value for $pp$ and $\bar{p}p$ scattering. The $\sigma_{gg}$ cross section, calculated via
expression (\ref{sloh1}), is shown in Fig. 3.

\begin{figure}
%\label{difdad}
%\vspace{2.0cm}
\begin{center}
%\vspace{-0.6cm}
\includegraphics[height=.29\textheight]{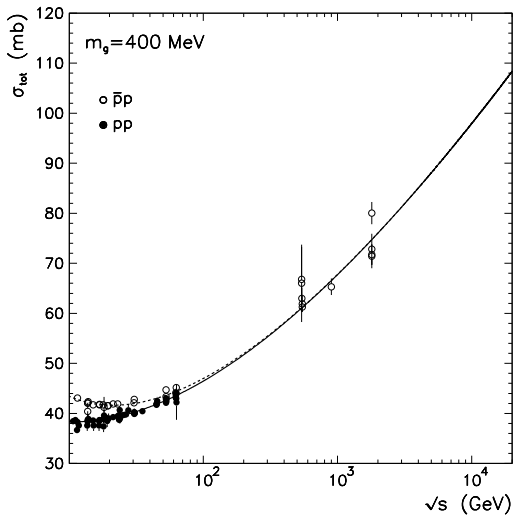}
\caption{Total cross section for $pp$ (solid curve) and $\bar{p}p$ (dashed curve) scattering.}
\end{center}
\end{figure}

\begin{figure}
%\label{difdad}
%\vspace{2.0cm}
\begin{center}
%\vspace{-0.6cm}
\includegraphics[height=.29\textheight]{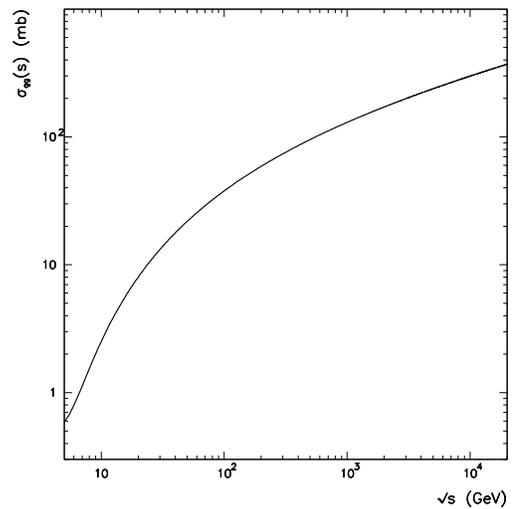}
\caption{Gluon-gluon total cross section. The dynamical gluon mass scale and the parameter
$C^{\prime}$ were set to $m_{g}=400$ MeV and to $C^{\prime}=(12.097\pm 0.962)\times 10^{-3}$, respectively.}
\end{center}
\end{figure}

\section{Photon-proton and photon-photon reactions}

Early modeling of hadron-hadron, photon-hadron and photon-photon  cross sections
within Regge theory shows a energy dependence similar to the ones of nucleon-nucleon
\cite{soft01,luna1bjp,luna1bjp2}. This
universal behavior, appropriately scaled in order to take into account the differences between hadrons and
the photon, can be understood as follows: at high center-of-mass energies the total
photoproduction $\sigma^{\gamma p}$ and the total hadronic cross section $\sigma^{\gamma \gamma}$
for the production of hadrons in the interaction of
one and two real photons, respectively, are expected to be dominated by
interactions where the photon has fluctuated into a
hadronic state. Therefore measuring the energy dependence of photon-induced processes should improve our
understanding of the hadronic nature of the photon as well as the universal high energy
behavior of total hadronic cross sections.

\begin{figure*}
\begin{center}
%\vspace{1.0cm}
\vglue 0.0cm
\hglue -8.0cm
\includegraphics[height=.28\textheight]{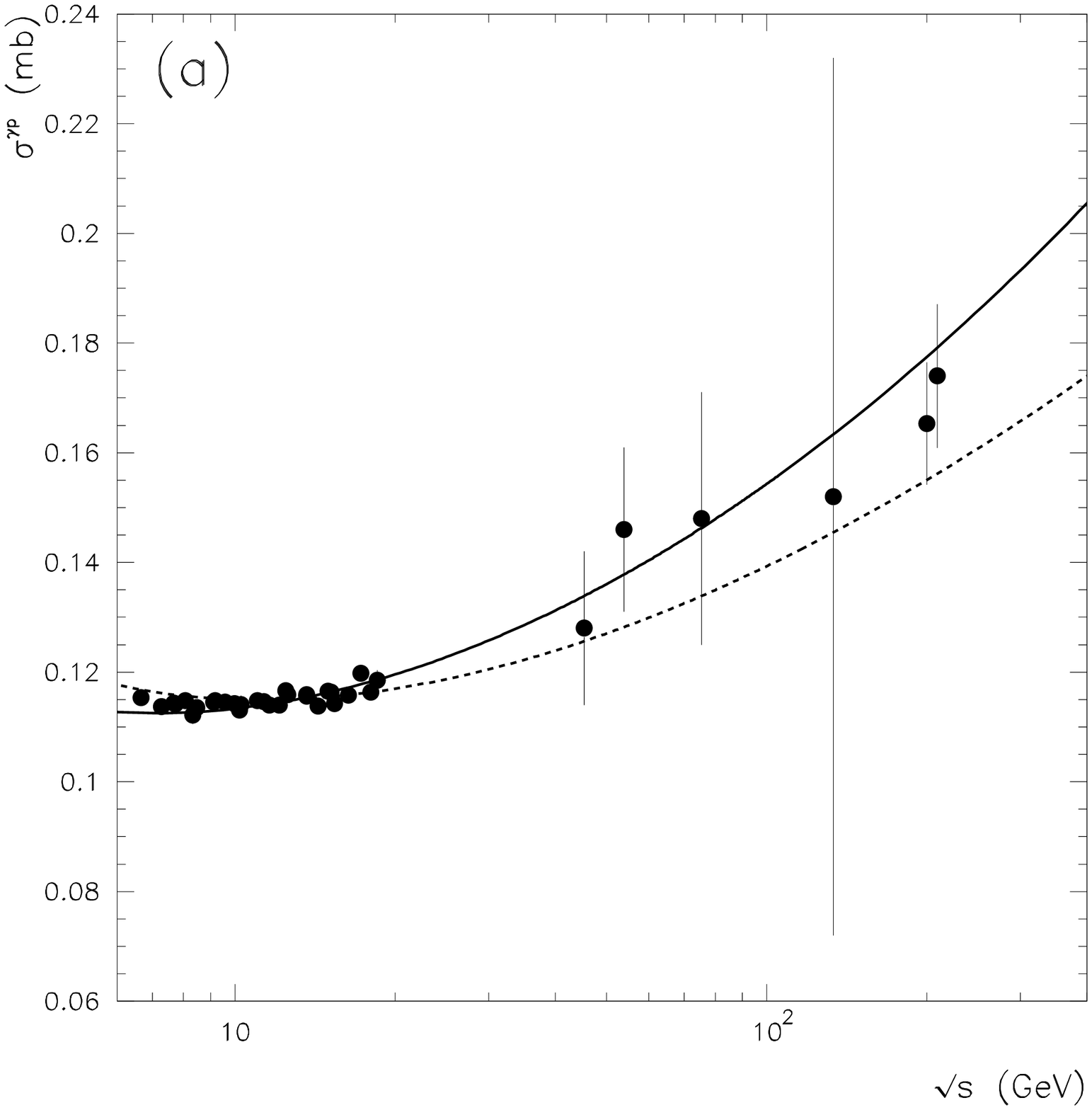}
\vglue -6.67cm
\hglue 8.3cm
\includegraphics[height=.28\textheight]{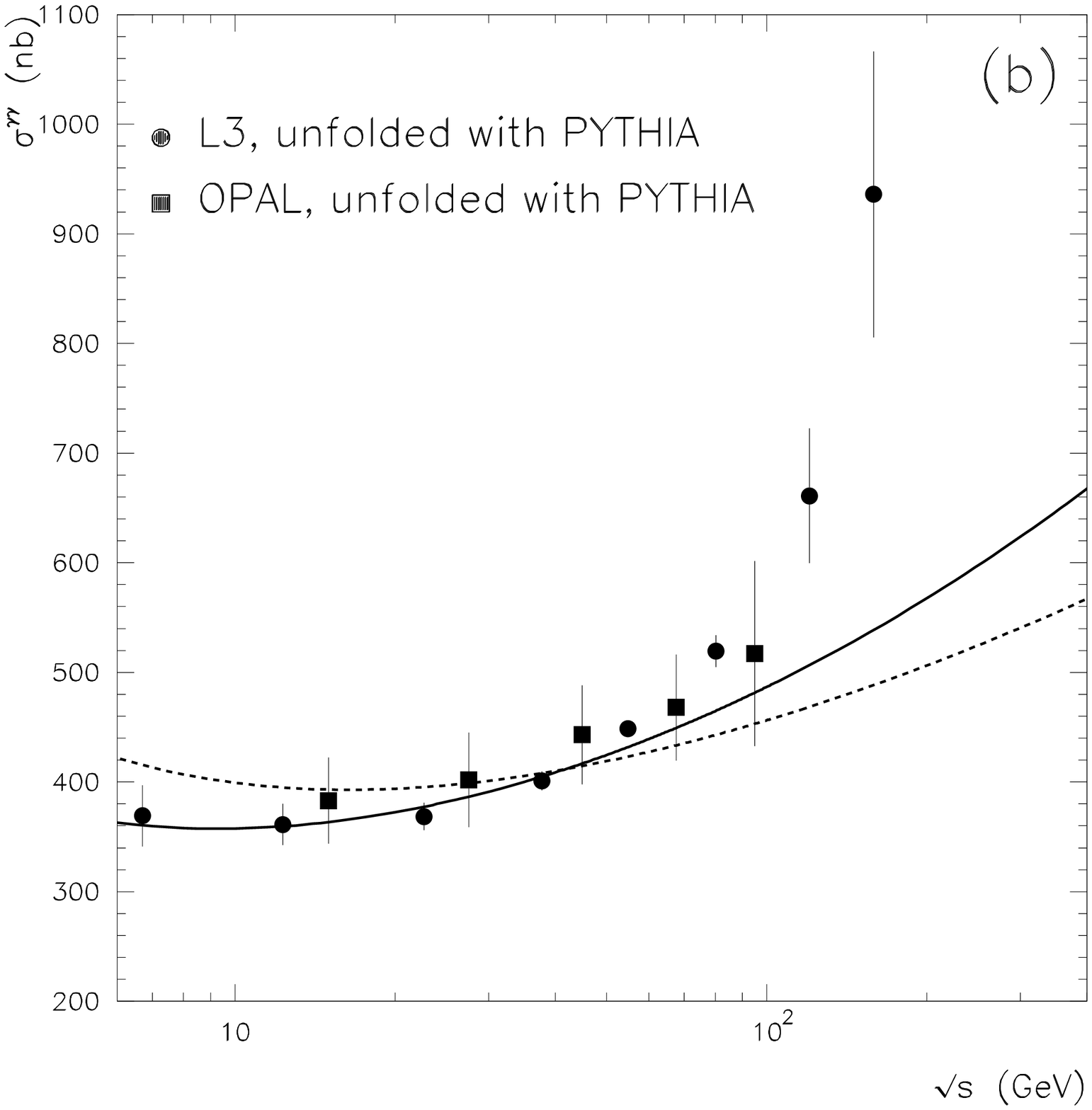}
\vglue 0.3cm
\hglue 8.3cm
\includegraphics[height=.28\textheight]{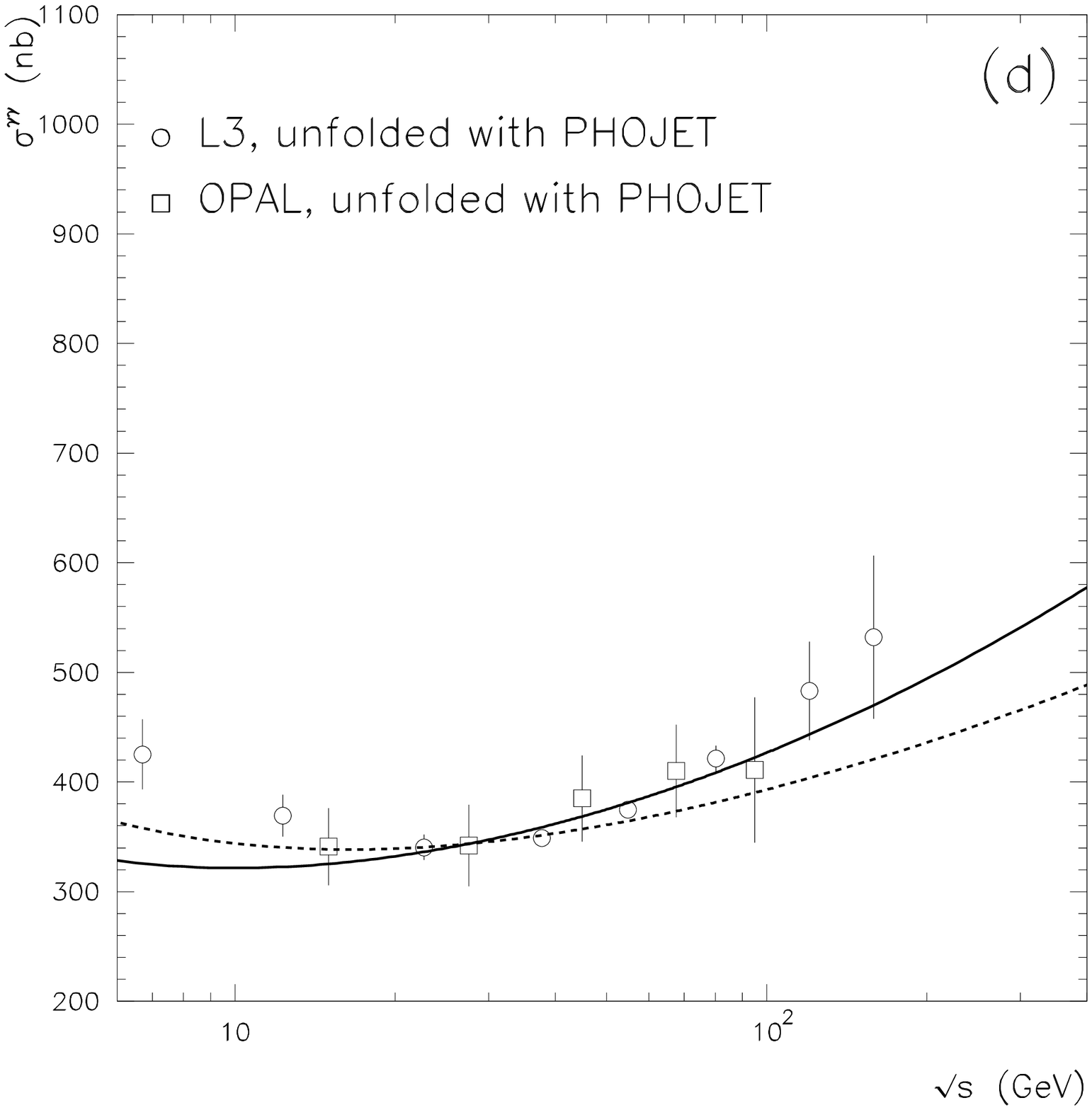}
\vglue -6.62cm
\hglue -8.0cm
\includegraphics[height=.28\textheight]{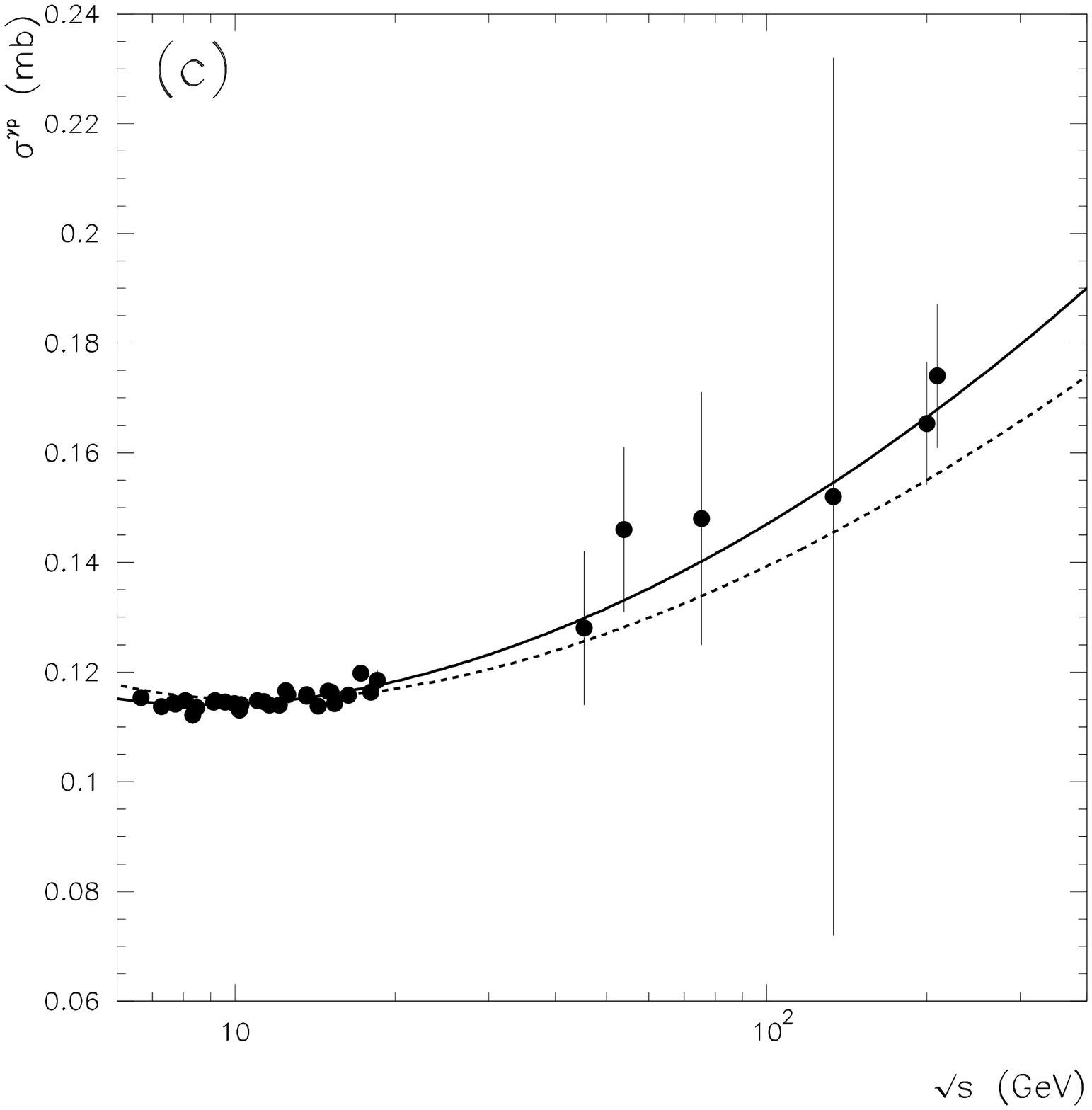}
\caption{$\sigma^{\gamma p}$ and $\sigma^{\gamma \gamma}$ cross sections corresponding to the case where
$P_{had}^{-1}$ varies with the energy (solid curves). The dashed curves correspond to the
case using a constant value of $P_{had}^{-1}$. The curves in (a) and (b) [(c) and (d)] are related to
the SET I [SET II].}
\end{center}
\label{frtd3}
\end{figure*}

However the comparison of the experimental data and the theoretical prediction
may present some subtleties depending on the Monte Carlo model used to analyze
the data. For example, the $\gamma \gamma$ cross sections are extracted from
a measurement of hadron production in $e^{+}e^{-}$ processes and are strongly dependent upon the acceptance
corrections to be employed. These corrections are in turn sensitive to the Monte Carlo models used in the
simulation of the different components of an event, and this general procedure
produces uncertainties in the determination of $\sigma^{\gamma \gamma}$ \cite{luna3bjp}. This clearly implies
that any phenomenological analysis has to take properly into account the discrepancies among
$\sigma^{\gamma \gamma}$ data obtained from different Monte Carlo generators.
Therefore we performed global fits considering separately data of the L3 \cite{acciarri} and OPAL
\cite{abbiendi} collaborations obtained through
the PYTHIA \cite{pythia} and PHOJET \cite{phojet} codes, defining two data sets as
\begin{center}
SET I: $ \sigma^{\gamma p}$ and $\sigma^{\gamma \gamma}_{PYT}$ data 
($\sqrt{s}_{\gamma p}, W_{\gamma \gamma} \geq 10$ GeV), \\
SET II: $ \sigma^{\gamma p}$ and $\sigma^{\gamma \gamma}_{PHO}$ data 
($\sqrt{s}_{\gamma p}, W_{\gamma \gamma} \geq 10$ GeV),
\end{center}
where $\sigma^{\gamma \gamma}_{PYT}$ ($\sigma^{\gamma \gamma}_{PHO}$) correspond
to the data of $\gamma \gamma$ total hadronic cross section obtained via the
PYTHIA (PHOJET) generator.

The even and odd amplitudes for $\gamma p$ scattering can be obtained
after the substitutions
$\sigma_{ij} \rightarrow (2/3)\, \sigma_{ij}$ and $\mu_{ij} \rightarrow \sqrt{3/2}\, \mu_{ij}$ in
the eikonals (\ref{final4}) and (\ref{oddeik}) \cite{luna3bjp}, where
$\chi^{\gamma \bar{p}}_{\gamma p}=\chi^{+}_{\gamma p}\pm \chi^{-}_{\gamma p}$. Assuming vector
meson dominance (VMD), the $\gamma p$ total cross section is given by
\begin{eqnarray}
\sigma^{\gamma p}(s) = 4\pi P_{had}^{\gamma p} \int_{_{0}}^{^{\infty}} \!\! b\, db\,
[1-e^{-\chi^{\gamma p}_{_{I}}(b,s)}\cos \chi^{\gamma p}_{_{R}}(b,s)] , \nonumber
\label{degt2}
\end{eqnarray}
where $P_{had}^{\gamma p}$ is the probability that the photon interacts as a hadron. In the simplest VMD
formulation this probability is expected to be of $\mathcal{O}(\alpha_{em})$:
\begin{eqnarray}
P_{had}^{\gamma p} = P_{had} = \sum \limits_{V=\rho, \omega, \phi} 
\frac{4\pi\alpha_{em}}{f^2_V} \sim \frac{1}{249} , \nonumber
\label{phvmd}
\end{eqnarray}
where $\rho$, $\omega$ and $\phi$ are vector mesons. However, there are expected contributions to $P_{had}$
other than $\rho$, $\omega$, $\phi$, as for example, of heavier vector
mesons and continuum states. Moreover, the probability $P_{had}$ may also depend on the energy, which is a
possibility that we explore in this work.

To extend the model to the $\gamma \gamma$ channel we just perform the substitutions  
$\sigma_{ij}\to (4/9)\, \sigma_{ij}$ and
$\mu_{ij}\to (3/2)\, \mu_{ij}$ in the even part of the eikonal (\ref{final4}). The calculation leads
to the following eikonalized total $\gamma \gamma$ hadronic cross section 
\begin{eqnarray}
\sigma^{\gamma \gamma}(s) = 4\pi N P_{had}^{\gamma \gamma}
\int_{_{0}}^{^{\infty}} \!\! b\, db\, [1-e^{-\chi^{\gamma
\gamma}_{_{I}}(b,s)}\cos \chi^{\gamma \gamma}_{_{R}}(b,s)] , \nonumber
\label{hudrt3}
\end{eqnarray}
where $P_{had}^{\gamma \gamma} = P_{had}^{2}$ and $N$ is a normalization factor which takes into account
the uncertainty in the extrapolation to real photons ($Q_{1}=Q_{2}=0$) of the hadronic cross section
$\sigma_{\gamma \gamma}(W_{\gamma \gamma},Q_{1}^{2},Q_{2}^{2})$ \cite{luna3bjp}.
With the eikonal parameters of the QCD eikonal model fixed by the $pp$ and $\bar{p}p$ data, 
we have performed all calculations of photoproduction and photon-photon scattering \cite{luna3bjp}. We have
assumed a phenomenological expression for $P_{had}$, implying that it increases logarithmically with the
square of the
center of mass energy: $P_{had} = a + b \ln (s)$. The total cross section
curves are depicted in Figure 3, where Figs. 3(a) and 3(b) [3(c) and 3(d)] are related to the SET I [SET II].
The results depicted in the Figures 3(c) and 3(d) show that the shape and
normalization of the curves are in good agreement with the data deconvoluted with PHOJET \cite{luna3bjp}.
The calculations using a constant value of $P_{had}$ (that does not depend on the energy $s$) are
represented by the dashed curves. These global results indicate that a energy dependence of $P_{had}$
is favored by the photoproduction and photon-photon scattering data.

\section{Conclusions}

In this work we have investigated the influence of an infrared dynamical gluon mass scale in the calculation
of $pp$, $\bar{p}p$, $\gamma p$ and $\gamma \gamma$ total cross sections through a QCD-inspired eikonal
model. By means of the dynamical perturbation theory (DPT) we have computed the tree level $gg\to gg$ cross
section taking into account the dynamical gluon mass. The connection between the subprocess cross section
$\hat{\sigma}_{gg}(\hat{s})$ and these total cross sections is made via a QCD-inspired eikonal model where
the onset of the dominance of gluons in the interaction of high energy hadrons is managed by the dynamical
gluon mass scale. By means of a global fit to the forward $pp$ and $\bar{p}p$ scattering data and to
$d\sigma^{\bar{p}p} /dt$
data at $\sqrt{s}=1.8$ TeV, we have determined the best phenomenological value of the dynamical gluon mass,
namely $m_{g}\approx 400^{+350}_{-100}$ MeV. Interestingly enough, this value
is of the same order of magnitude as
the value $m_{g}\approx 500 \pm 200$ MeV, obtained in other calculations of strongly interacting processes.
This result corroborates theoretical analysis taking into account
the possibility of dynamical mass generation and show that, in principle, a dynamical nonperturbative
gluon propagator may be used in
calculations as if it were a usual (derived from Feynman rules) gluon propagator. 

With the help of vector meson dominance and the additive quark model, the QCD model can successfully describe
the data of the total photoproduction $\gamma p$ and total hadronic $\gamma \gamma$ cross sections. We have
assumed that $P_{had}$ has a logarithmic increase with $s$. This choice
leads to a improvement of the global fits, i. e. the logarithmic increase of $P_{had}$ with $s$ is quite
favored by the data. Notice that the data of $\sigma^{\gamma \gamma}_{PYT}$ above
$\sqrt{s} \sim 100$ GeV can hardly be described by the QCD model. Assuming
the correctness of the model we could say that the PHOJET generator is
more appropriate to obtain the $\sigma^{\gamma \gamma}$ data above 
$\sqrt{s} \sim 100$ GeV. This conclusion is supported by the recent result that the factorization 
relation does not depend on the assumption of an 
additive quark model, but more on the opacity of the
eikonal being independent of the nature of the reaction \cite{kaidalov}.

\noindent {\bf Acknowledgments}:
I am pleased to dedicate this paper to Prof. Yogiro Hama, on the occasion of his
70th birthday. I am grateful to the editors of the Braz.
J. Phys. who gave me the opportunity of contributing to the volume in his honor, and to
M.J. Menon and A.A. Natale for useful comments. This research was supported by the
Conselho Nacional de Desenvolvimento
Cient\'{\i}fico e Tecnol\'ogico-CNPq under contract 151360/2004-9.
%\end{acknowledgments}

\end{document}